\documentclass[a4paper,11pt]{article}

\pdfoutput=1 
\usepackage{jcappub} 
\usepackage{eurosym}

\usepackage[T1]{fontenc} 

\usepackage{amssymb}
\usepackage{xcolor}
\usepackage[normalem]{ulem}
\usepackage{systeme}

\newcommand{\be}{\begin{equation}}
\newcommand{\ee}{\end{equation}}
\newcommand{\bea}{\begin{eqnarray}}
\newcommand{\eea}{\end{eqnarray}}
\newcommand{\bfig}{\begin{figure}}
\newcommand{\efig}{\end{figure}}
\newcommand{\nn}{\nonumber}

\newcommand{\ylm}[1]{Y_{\ell m}( #1)}

\def\d{{\rm d}}
\def\H{{\mathcal H}}
\def\n{\boldsymbol{n}}
\def\k{\boldsymbol{k}}
\def\v{\boldsymbol{v}}
\def\alm{a_{\ell m}}
\def\l{\left(}
\def\r{\right)}
\def\D{\Delta}
\def\p{\partial}
\def\fsky{f_{\rm sky}}

\def\odm{{\Omega_{\rm cdm  0}}}
\def\ob{{\Omega_{\rm b 0}}}

\def\cs{{\sc CAMB$\_$sources\,}}

\title{Constraints on the growth rate using the observed galaxy power spectrum}

\author[1,2,3]{Jos\'e Fonseca,}
\author[3]{Jan-Albert Viljoen,}
\author[3,4]{Roy Maartens\,}
\affiliation[1]{Dipartimento di Fisica e Astronomia ``G. Galilei'', Universit\`{a} degli Studi di Padova, Via Marzolo 8, 35131 Padova, Italy}
\affiliation[2]{INFN, Sezione di Padova, via Marzolo 8, I-35131, Padova, Italy}
\affiliation[3]{Department of Physics \& Astronomy, University of the Western Cape, Cape Town 7535, South Africa}
\affiliation[4]{Institute of Cosmology \& Gravitation, University of Portsmouth, Portsmouth PO1 3FX, UK}

\emailAdd{josecarlos.s.fonseca@gmail.com}

\abstract{The large-scale structure growth index $\gamma$ provides a consistency test of the standard cosmology and is a potential indicator of modified gravity. We investigate the constraints on $\gamma$ from next-generation spectroscopic surveys, using the power spectrum that is observed in redshift space, i.e., the angular power spectrum. The angular power spectrum avoids the need for an Alcock-Packzynski correction. It also naturally incorporates cosmic evolution and wide-angle effects, without any approximation.  We include the cross-correlations between redshift bins, using a hybrid approximation when the total number of bins is computationally unfeasible.  We show that the signal-to-noise on  $\gamma$ increases as the redshift bin-width is decreased. Noise per bin also increases -- but this is compensated by the increased number of auto- and cross-spectra. In our forecasts, we marginalise over the amplitude of primordial fluctuations and other standard cosmological parameters, including the dark energy equation of state parameter, as well as the clustering bias. Neglecting cross-bin correlations increases the  errors by $\sim40 - 150\%$. Using only linear scales, we find that  a DESI-like BGS  survey and an HI intensity mapping survey with the SKA1 precursor MeerKAT deliver similar errors of $\sim4-6\%$, while a Euclid-like survey and an SKA1 intensity mapping survey give $\sim3\%$ errors. }

\begin{document}
\maketitle
\flushbottom

\section{Introduction}

We are entering a new era in the study of the large-scale structure (LSS) of the Universe. Not only will we map the sky over larger areas, but we will also go deeper in redshift. As well as the increasing volumes, we will probe the sky in different frequency ranges, creating complementary sets of dark matter tracers. 

Einstein's theory of gravity and its modifications (see e.g. the reviews  \cite{Clifton:2011jh,Koyama:2015vza,Langlois:2018dxi}) leave distinctive imprints on the clustering of matter and its peculiar velocity. Identifying the statistical effect of peculiar velocities on the distribution of matter provides a powerful test of the cosmological model and the theory of gravity. This test is based on using redshift-space distortions (RSD) to measure the LSS growth rate $f$ or  growth index $\gamma=\ln f/ \ln \Omega_{\rm  m}$. To implement this test one requires the redshift accuracy of spectroscopic surveys. Upcoming spectroscopic surveys, in optical, near infra-red and  in  radio bands \cite{Laureijs:2011gra,chime,2012IJMPS..12..256C,2012arXiv1209.1041B,Levi:2013gra,Dore:2014cca,Newburgh:2016mwi,Santos:2017aa,Bacon:2018dui}, will have higher redshift accuracy and cover larger sky areas than ever,  allowing for higher precision tests. 

The standard analysis of RSD data for tests of gravity uses the spatial power spectrum $P_g(\k,z)$ in Fourier space, which allows one to cleanly separate the RSD effect via a Legendre multipole expansion (see e.g. \cite{Zhao:2018jxv} for the current state of the art). Similarly, most forecasts for future surveys rely on the same analysis (e.g. \cite{Bull:2015lja,Amendola:2016saw,Pourtsidou:2016dzn}). The Fourier power spectrum requires a choice of fiducial model to convert observed angles and redshifts to distances. This then requires an  Alcock-Paczynski correction to compensate for the error in the choice of fiducial. In addition, the standard Fourier analysis implicitly neglects cosmic evolution in the rather thick redshift bins used, and it encodes a flat-sky approximation that neglects wide-angle correlations.  There are prescriptions to deal with these issues (e.g.  \cite{Ruggeri:2017rza}). However, it is useful to explore an alternative analysis that avoids these issues from the start (while of course introducing other issues).

 The alternative is to use the angular power spectrum $C_\ell(z,z')$, which is the harmonic transform of the correlation function that is observed in redshift space -- i.e., on the backward light-cone of the observer~\cite{Challinor:2011bk, Bonvin:2011bg,Bruni:2011ta, DiDio:2013sea, Tansella:2017rpi}. The angular power spectrum of the observed data does not require a fiducial model and therefore does not need an Alcock-Paczynski correction. Furthermore, it naturally incorporates wide-angle correlations and cosmic evolution.  It also naturally incorporates Doppler and lensing magnification effects on the correlations, which we include, as well as other smaller relativistic observational effects, which we neglect here.
  
An immediate issue with $C_\ell(z,z')$ is that, unlike $P_g(\k,z)$, we cannot cleanly separate out the RSD effect. In addition, computational  complexity arises from the oscillating spherical Bessel functions in $C_\ell(z,z')$, from cross-bin correlations for $z'\neq z$, and from the need for very thin redshift bins to maximally exploit the potential of spectroscopic surveys. Nevertheless, advances in using the angular power spectrum to analyse galaxy survey data are ongoing (e.g. \cite{Asorey:2012rd,Campagne:2017xps,Assassi:2017lea,Gebhardt:2017chz,Camera:2018jys,Schoneberg:2018fis,Loureiro:2018qva, Alonso:2018jzx,Tanidis:2019teo}). 

It is intuitively clear that thick redshift bins in $C_\ell$ will wash out the RSD signal so that thin bins are needed for precision on RSD. As redshift bins are decreased in size, the problem of noise grows, so that one might expect to reach a `sweet spot' in bin width where the precision is optimal. This expectation is however not correct, since it ignores the additional information that arises from the growing number of auto- and cross-correlations. We confirm that information on the growth index $\gamma$ continues to increase  with decreasing bin width, reaching  a theoretical maximum for infinitely thin bins. In practice we need to choose a bin width that is feasible for numerical computation.

We use Fisher forecasting and marginalise over the standard cosmological parameters, in particular including the amplitude of primordial fluctuations and the dark energy equation of state parameter, as well as over the clustering bias {in each bin}. Modelling nonlinear RSD is beyond the scope of this work and therefore we use information only from linear scales.

We find that the errors on  $\gamma$ for a Euclid-like survey and an SKA1 neutral hydrogen (HI) intensity mapping survey are $\sim3\%$. We also find that an HI  intensity mapping survey with the SKA1 precursor MeerKAT and a DESI-like BGS survey have a similar accuracy of $\sim 5\%$. Percent-level errors seem to be only within reach for the futuristic SKA2 HI galaxy survey. We also show that if only auto-correlations $C_\ell(z,z)$ are used and cross-bin correlations $C_\ell(z,z')$ are neglected,  then the constraints degrade by a factor of $\sim 40-150\%$.
 
This paper is structured as follows. In \S \ref{sec:modgrav} we review the effect of peculiar velocities in the angular power spectrum.  \S \ref{sec:surveys} describes the technical assumptions on the surveys and tracers that we consider. In \S \ref{sec:optbin} we analyse the effect of redshift binning  on RSD precision. Our results are  presented in \S \ref{sec:results}, and we conclude in \S \ref{sec:conclusion}. Our fiducial model is a concordance LCDM model with the Planck  2018 best-fit parameters.

\section{Constraining the growth rate using the angular power spectrum} \label{sec:modgrav}

On linear scales,  peculiar velocities are sourced by matter density gradients via the continuity equation,
\bea \label{eq:linvel}
\nabla\cdot \v=f{\H}\, \delta^{\rm c}\,,
\eea
where ${\cal H}=(\ln a)'$ is the conformal Hubble rate and $\delta^{\rm c}$ is the comoving matter density contrast. The growth rate is 
\be
f \equiv \frac{\partial \ln D}{\partial \ln a}\,,
\ee 
where $D$ is the growth factor. The growth rate at any redshift $z$ is clearly affected by the relative amount of dark matter at $z$, and we can alternatively use the growth index $\gamma$, defined by 
\be \label{eq:f_gamma}
f = \big(\Omega_{\rm m}\big)^\gamma \,.
\ee
For the standard LCDM model, $\Omega_{\rm m}=\Omega_{{\rm m} 0}(1+z)^3[\Omega_{{\rm m} 0}(1+z)^3+1-\Omega_{{\rm m} 0}]^{-1}$,  and
 $\gamma =0.545$ is a very good approximation. This approximation remains good for simple models of dynamical dark energy. For modified gravity models, the approximation breaks down and $\gamma$ is significantly different.
 Hence measuring the growth index (or the growth rate) provides a test of LCDM and potentially of General Relativity. Note that $f$ and $\gamma$ are scale-independent in LCDM (and to a good approximation in simple extensions of LCDM), and the redshift dependence of $f$ is determined by $\Omega_{\rm m}$.
In modified gravity models there is typically scale dependence in $f$ and thus in $\gamma$, and $\gamma$ will also be redshift dependent. Here our focus is on a new approach to growth constraints via the angular power spectrum and we assume constant $\gamma$, which applies to LCDM and wCDM, the simple extension that we consider.

The number of sources  counted by the observer in a solid angle element about unit direction $\n$ and in a redshift interval is given by 
\be
\d \mathbb{N} =N \d z\,\d\Omega_{\n}=  {\cal N}\, \d V \,. 
\ee
Here $N$ is the number that is counted by the observer per redshift per solid angle. By contrast, 
$  {\cal N}$ is the proper number density, which is not observed by the observer but is the quantity that would be measured at the source. Similarly,  $\d V$ is not the observed volume element but the proper volume element corresponding to $\d z$ and $\d\Omega_{\n}$, as measured at the source.  Then the observed number density contrast is $\delta_N$, which is related to the proper number density contrast at the source as  
\bea
 \delta_N&=&\delta_{ {\cal N}} +\,\mbox{RSD + Doppler +  lensing corrections} \notag\\
\label{obdel}
&=& \delta_{ {\cal N}}-{1\over {\cal H}}\n \cdot \nabla \big(\v  \cdot \n \big)  +  A\big(\v  \cdot \n \big)+ (2-5s)\kappa\,,
\eea
where  $A$ and $\kappa$  are given in harmonic space in \eqref{eq:fulltransfer} below and $s$ is given in (\ref{eq:sg}). The observed two-point correlation function then defines the angular power spectrum:
\be
\big\langle  \delta_N(z_i,\n)\,  \delta_N(z_j,\n')\big\rangle = \sum_{\ell}\,{(2\ell+1) \over 4\pi}\, C_\ell(z_i,z_j)\,{\cal L}_\ell \big(\n\cdot\n'\big)\,,
\ee
where ${\cal L}_\ell$ are Legendre polynomials.

The number density contrast observed on the backward light-cone can be given directly in terms of the  fundamental  observables $z_i$ and $\n$ by expanding  in spherical harmonics:
\be
\delta_N(z_i, \n)= \sum_{\ell m}  \alm(z_i)\, \ylm{\n}\,.
\ee
For now we assume that the redshift bin is an infinitesimal shell centred at $z_i$, and then we treat the realistic case below. By statistical homogeneity and isotropy, $\langle a_{\ell m} \rangle = 0$, and we can use the $\alm$ and their covariance as estimators of the cosmological parameters. We assume that the $\alm$ are Gaussian distributed with covariance
\be
\big\langle a_{\ell m}(z_i)\, a^{*}_{\ell' m'}(z_j)\big\rangle = \delta_{\ell \ell'}\, \delta_{m m'}\, C_\ell(z_i,z_j)\,.
\ee

\subsection{{Angular transfer functions}}

We follow the notation and approach of \cite{Challinor:2011bk} and relate the covariance of the $\alm$ with the primordial power spectrum of the curvature perturbation $\mathcal P (k)$ and the theoretical transfer functions $\D_\ell$ as 
\be \label{eq:clgeneral}
C_\ell \l z_i,z_j \r=4\pi\!\!\int\!\!\d \ln k\, \D_\ell\l z_i,k\r \D_\ell\l z_j,k\r\, \mathcal P (k).
\ee
The primordial power spectrum $\mathcal P (k)=A_s\l k/k_0\r^{n_s-1}$ encodes information about the seeds of structure formation via its amplitude $A_s$ and spectral index $n_s$. The pivot scale is set to $k_0=0.05\,$Mpc$^{-1}$. The transfer functions include the tracer density contrast, as well as RSD and all other observational effects on the 2-point correlation function. In this paper we focus on the effects from  peculiar velocities (RSD and the Doppler effect) and also include the effect of lensing magnification. Further effects are suppressed by a factor ${\cal H}^2/k^2$ and we  neglect  them. 
(The full expression can be found in \citep{Challinor:2011bk,Bonvin:2011bg,Bruni:2011ta,Jeong:2011as} for galaxy surveys and in \citep{Hall:2012wd,Fonseca:2018hsu} for maps of intensity.) 
Then the theoretical transfer function  is given by 
\bea
\D_\ell(k)&=& b\,\delta^{\rm c}_k\,
j_\ell\l k\chi\r+\frac {k v_k}{\H}\,j_\ell''(k\chi)\nn\\
&&{}+\l\frac{2-5s}{\H\chi} +5s-b_e+\frac{\H'}{\H^2}\r  v_k j_\ell'(k\chi)  \nn\\
&&{}+\frac{\ell\l\ell+1\r\l2-5 s\r}2\int_0^{\chi}
\d\tilde\chi\frac{(\chi-\tilde\chi)}{\chi\tilde\chi}  \big[\phi_k(\tilde\chi)+\psi_k(\tilde\chi)\big] j_\ell\l k\tilde\chi\r
\,, \label{eq:fulltransfer}
\eea
where we suppressed the redshift dependence to simplify notation. The first line corresponds to the standard density plus RSD terms, the second line is the  Doppler term and the third line the  lensing contribution. Here $\chi$ is the comoving line-of-sight distance, $j_\ell$ are spherical Bessel functions and the perturbed metric is given in Poisson gauge by
\bea 
{\d}s^2=a^2\ \Big[-\big(1+2\psi\big)\, {\d}\eta^2 + \big(1-2\phi\big)\, {\d}\boldsymbol{x}^2\Big].
\eea
In addition to the clustering bias $b$ in \eqref{eq:fulltransfer}, we also have the `evolution bias',  accounting for redshift evolution of sources, and the magnification bias, accounting for the way that lensing alters the  number of sources that actually enter in the survey:
 \bea \label{eq:be}
b_e&=& \frac{\p\ln a^3 \bar{\cal N}}{\p\ln a}\,, \\
s&=&\frac{\p \log_{10}  \bar{\cal N}} {\p m_\ast}\,. \label{eq:sg}
\eea
Here $m_\ast$ is the threshold magnitude of the survey and the background number density of sources is $\bar{\cal N}=\bar{\cal N}( z,m<m_*)$. In the background, $\bar N=\bar{\cal N} \chi^2(1+z)^{-4}{\cal H}^{-1}$.
(From now on we drop the overbar on the background values of ${\cal N}, N, \mathbb{N}$.)

So far we assumed that the survey  window function is given by a delta function. In reality, the observed transfer function $\D^{W}_\ell(z_i,k)$ includes the fact that the window function $W$ may weight different redshifts differently and that the redshift distribution of sources $p$ may not be constant in redshift:
\bea \label{eq:obs_tranf}
\D_\ell^{W}( z_i,k)=\int\!\! \d z\,p(z)\ W(z_i,z)\ \D_\ell(z,k)\,.
\eea
Hence $p$ and $W$ work as redshift weighting functions of the theoretical transfer function. Note that the product $p W$ is thus normalised to unity {for galaxy surveys}: $\int\!\d z\,p(z)W(z_i,z)=1$ for all $z_i$. In fact $p\propto N=\d\mathbb{N}/(\d z\d\Omega)$, the (background) observed number density per $z$ per $\Omega$ for galaxy surveys (see below for the case of intensity mapping) so we will present the window function in the next sections leaving all normalisations included in $p$. 

\subsection{{Fisher matrix analysis}}

We write the Fisher matrix for a set of parameters $\vartheta_\alpha$ as \citep{Tegmark:1996bz}
\bea \label{eq:fishercl}
{ F}_{\vartheta_\alpha \vartheta_\beta}=\sum_{\ell_{\rm min}}^{\ell_{\rm max}} \frac{(2\ell+1)}{2 }\fsky\,{\rm Tr}\Big[ \big(\partial_{\vartheta_\alpha}{  C}_{\ell}\big){ \Gamma}_\ell^{-1} \big(\partial_{\vartheta_\beta}{  C}_{\ell}\big){ \Gamma}_\ell^{-1}\Big]\,,
\eea
{where $C_\ell$ is the matrix $C_\ell^{ij}=C_\ell(z_i,z_j)$.}
The survey sky fraction $\fsky=\Omega_{\rm survey}/ 4\pi$ is an approximation accounting for the $m$ summation. For simplicity we have not binned in $\ell$. The observed covariance includes the noise term, $\Gamma_\ell=C_\ell+{\rm Noise}_\ell$, and we assume that the noise terms do not depend on the cosmological parameters. 
In practice one uses $C_\ell$ and its covariance (see~\cite{DiDio:2013sea} for details) as the observable at high $\ell$ {where we can safely assume that its likelihood is Gaussian} and the $\alm$ only at low $\ell$ {where it is computationally feasible}. For Fisher forecasts, the two approaches are equivalent.
 
The lower limit of the sum in \eqref{eq:fishercl}  is the largest angular scale in the survey. If a survey covers a single  spherical cap, then the maximum angular scale available is given by the survey area, i.e., 
\be
\ell_{\rm min}={\rm int}\Big(\pi/\sqrt{\Omega_{\rm survey}}\Big)+1\,. 
\ee
To allow for more realistic sky coverage, we use a smaller $\ell_{\rm min}$. Furthermore,
in the case of HI intensity mapping, the removal of foregrounds affects the largest scales, $\ell_{\rm min} \lesssim 5$~\cite{Witzemann:2018cdx,Cunnington:2019lvb}. We impose $\ell_{\rm min}=5$ in all surveys. Scales with $\ell_{\rm min}<5$ will contribute negligibly to constraints on the growth rate. 

Under the assumption that the $\alm$ likelihood is Gaussian, the inverse of the Fisher matrix is a good approximation of the parameter covariance. Hence the forecasted marginal and conditional errors for a parameter $\vartheta_\alpha$ are given by
\bea
\sigma_{\vartheta_\alpha}=\left [({  F}^{-1})_{\vartheta_\alpha\vartheta_\alpha}\right]^{1/2}\,,\qquad\qquad
\sigma^{\rm cond}_{\vartheta_\alpha}=\l F_{\vartheta_\alpha\vartheta_\alpha}\r ^{-1/2}\,.
\eea

We consider the following set of parameters:
\be\label{para}
  \vartheta_\alpha = \big\{ \ln \gamma, \ln A_s, b(z_i), \ln n_s,\ln\odm,\ln \ob, w, \ln H_0   \big\}.
\ee  
For the cosmological parameters we use the fiducial values: $A_s=2.142\times 10^{-9}$, $n_s=0.967$, $\odm=0.26$, $\ob=0.05$, $w=-1$, $H_0=67.74\,$ km/s/Mpc. For the growth index, we take $\gamma=0.545$, as discussed above. In addition, the clustering bias in each bin, $b(z_i)$, is a free parameter, with fiducial value set by the  bias models for each survey (see \S \ref{sec:surveys}). We assumed Gaussian   priors from Planck 2018 \cite{Aghanim:2018eyx} for all cosmological parameters. The priors are on the parameters, not their logarithms; we use the log of the parameters in the Fisher matrix purely because this gives a numerically more stable inversion of the matrix (see \cite{Camera:2018jys} for more detail).

The number of bias parameters is survey- and binning-dependent and is discussed in \S \ref{sec:surveys}. As will become clear, in some experiments the number of bins introduces computational limitations, and we develop a means to deal with this, following the idea proposed in \cite{Camera:2018jys}.

The angular power spectra and their derivatives are computed using \cs\!. The derivatives with respect to $\ln\odm,\ln \ob,w, \ln H_0$ are taken numerically using the 5-point stencil method. For the remaining parameters, we use a modified \cs to accept analytical derivatives.\footnote{\url{https://github.com/ZeFon/CAMB\_sources\_MT\_ZF.git}} 
The analytical derivative with respect to $\gamma$ uses the parametrisation \eqref{eq:f_gamma}, so that $\p f/\p \gamma=f\, \ln \Omega_{\rm m}$. Then  
\bea
\frac{\partial }{\partial \gamma}C_\ell(z_i,z_j)&=& \int {\rm d}\ln k\  \bigg[ \Delta^{W}_\ell(z_j) \int {\rm d}z'\ p(z')\ W(z',z_i)\ \ln \Omega_{\rm m}\ \Delta^{\v}_\ell \nn \\
&&{}+\Delta^{W}_\ell(z_i) \int {\rm d}z''\ p(z'')\ W(z'',z_j)\ \ln \Omega_{\rm m}\ \Delta^{\v}_\ell \bigg]\ {\cal P}(k)\,,
\eea
where $\Delta^{\v}_\ell$ denotes the RSD and Doppler terms in \eqref{eq:fulltransfer}. 
The analytical derivatives with respect to $A_s$, $n_s$ and the biases are given in Appendix A of \cite{Fonseca:2018hsu}. 

\subsection{{Excluding nonlinear effects}}

\begin{figure}[!h]
\centering
\includegraphics[width=7cm]{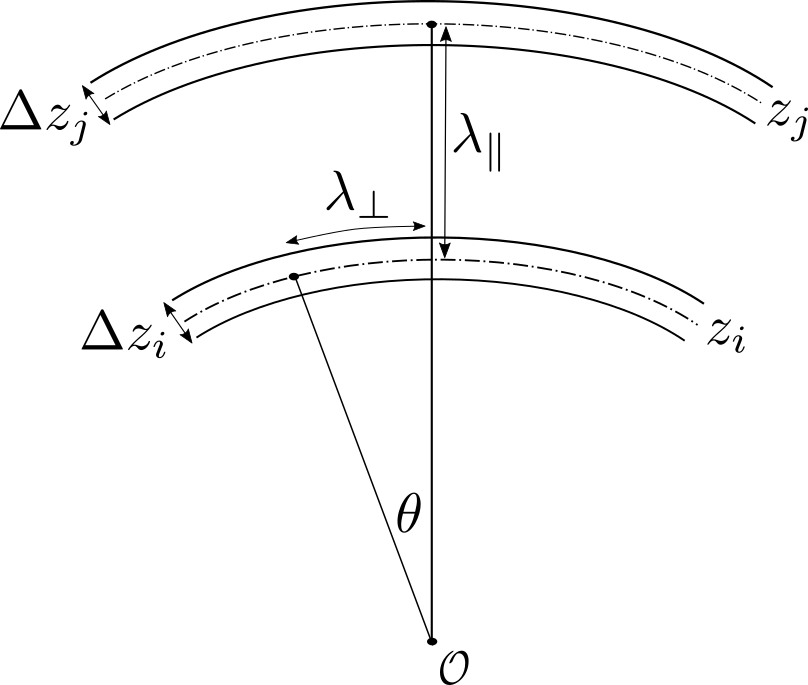}
\caption{Scales associated with a multipole $\ell$. }
\label{fig:diag_NL_scale}
\end{figure}

When computing  forecasts we only consider linear perturbations. The inclusion of nonlinear effects is beyond the scope of this paper and we therefore need to choose $\ell_{\rm max}$ in \eqref{eq:fishercl} to exclude these effects from the Fisher matrix. In Fourier space, the  scale marking the breakdown of a perturbative analysis of matter clustering is given in \citep{Smith:2002dz} as
\be
k_{\rm nl}(z)=k_{\rm nl,0}\l1+z\r^{2/(2+n_s)} \quad\mbox{with}\quad k_{\rm nl,0}\simeq 0.2 h\,{\rm Mpc}^{-1} \,.
\ee
On the scales of interest, the multipole $\ell$ corresponds to  a transverse comoving length scale $\lambda_\perp=2\pi/k$, which subtends an angle $\theta=\lambda_\perp/\chi$ at the observer (see Figure \ref{fig:diag_NL_scale}).
Using $\theta=2\pi/\ell$, this gives $\ell= \chi k$. It follows that, for auto-correlations, the
maximum multipole is
\be
\ell_{\rm max}(z_i,z_i) = \chi(z_i) k_{\rm nl}(z_i)\,.
\ee

The case of cross-bin correlations is more complicated (see the discussion in~\cite{DiDio:2013sea}). 
In principle, one can allow for $\ell\gg 1$  in the case of near-radial correlations, {for which $\theta \ll 1$ and $\lambda_\perp \ll 2\pi/k_{\rm nl}$ -- provided that $\lambda_\| > 2\pi/k_{\rm nl}$, to ensure that $\lambda > 2\pi/k_{\rm nl}$.}  However, 
in order to fully exclude nonlinearities from the Fisher matrix, we need to exclude them also from the covariance of $C_\ell^{ij}$, which contains the term $C_\ell^{ii}C_\ell^{jj}$. If $\lambda_\perp$ is nonlinear then $C_\ell^{ii}$  includes nonlinear  effects   and hence the covariance contains nonlinearities. Therefore we impose the cut
\be \label{lmax}
\ell_{{\rm max}}(z_i,z_j)={\rm min}\big\{\ell_{{\rm max}}(z_i,z_i),\ell_{{\rm max}}(z_j,z_j)\big\}\,.
\ee

\section{Future spectroscopic surveys}\label{sec:surveys}

To access the information encoded in the velocity field one requires the high redshift resolution of spectroscopic galaxy count surveys or intensity mapping (IM) surveys. We focus only on HI IM; in the future other line IM surveys may be available. 

In galaxy surveys the main source of uncertainty is the shot noise:
\be \label{slg}
{\cal S}^g_\ell(z_i,z_j)  =\frac1{ {N_\Omega} (z_i)}\, \delta_{ij}\,,
\ee
where  the average angular density of sources in the bin is
\be \label{eq:ngalbin}
 {N_\Omega} (z_i)=\int \d z\, W(z,z_i)\,  N(z)\,.
\ee 
For photometric galaxy surveys the window function is generically an error function \cite{Ma:2005rc}, which one can extend to spectroscopic redshift surveys but with narrow redshift uncertainties. For spectroscopic surveys this error function window function becomes very close to a smooth top-hat, which {is also} 
numerically stable. We use the smoothed top-hat window:
\be \label{eq:Wths}
\!\!  W(z,z_i;{\Delta} z_i,{\sigma_z})\!=\!\frac1{2\,{\tanh}({\Delta} z_i/2\sigma_{zi})} \left[  \tanh\!\l \!\frac{z-z_i+{\Delta} z_i/2}{\sigma_{zi}}\!\r\! -\! \tanh\!\l \!\frac{z-z_i-{\Delta} z_i/2}{\sigma_{zi}}\! \r\! \right]\!,
\ee
where $\sigma_z$ is the redshift resolution  and $\sigma_{zi} =\sigma_z(1+z_i)$.
This window function applies to both spectroscopic galaxy and IM surveys. 

In the case of HI IM, the noise component is thermal, coming mainly from the instrument. For single-dish IM experiments, where the dish signals are simply added rather than combined interferometrically, the instrumental noise is similar to CMB experiments \cite{Knox:1995dq}, and is given by
\be \label{eq:inst_noise_hiim}
{\cal T}^{\rm HI}_\ell(z_i,z_j)  =\frac{4\pi\,f_{\rm sky}\, T_{\rm sys}^2}{2\, N_{\rm d}\, \Delta \nu\, t_{\rm tot}}
\, \delta_{ij}\,.
\ee 
Here  $T_{\rm sys}$ is the system temperature of the telescope (including contributions from the receiver and the sky), $N_{\rm d}$ is the number of dishes, $\Delta \nu$ is the band size and $t_{\rm tot}$ is the total integration time. The factor of $2$ comes from the two polarisations. 

The angular power spectrum  $C^{\rm HI}_\ell$ of the theoretically observed HI temperature fluctuations is modified by the effect of the telescope beam:
\be
C^{\rm HI}_\ell(z_i,z_j) ~\rightarrow~ C^{\rm beam,HI}_\ell \l z_i,z_j\r=B_\ell\l z_i\r  B_\ell\l z_j\r C^{\rm HI}_\ell \l z_i,z_j\r\,.
\ee
Assuming a Gaussian beam, we have 
\be
B_\ell=\exp\Big[-\frac{\ell\l\ell+1\r}{16\ln 2}\,\theta_{\rm res}^2\Big]\,,
\ee
where the angular resolution of the dishes with diameter $D_{\rm d}$ is 
\be
\theta_{\rm res}=1.22\, \frac{\lambda_{\rm obs}}{D_{\rm d}} ~ \rm rad\,.
\ee

\subsection{HI intensity mapping surveys}

Intensity mapping of the 21cm HI emission line (after reionization) integrates the total emission from all galaxies in a pixel, giving a map of fluctuations at each $z$, with extremely small redshift error. This includes the emission from the brightest galaxies as well as from otherwise undetectable faint objects. The properties of this tracer are survey independent, which is not the case for galaxy surveys. We follow \cite{Santos:2015bsa} and use a halo-based model for the HI average temperature and clustering bias, with the fitting  functions:
\bea
\bar{T}^{\rm HI}(z)&=&0.056 + 0.232 z - 0.024 z^2\,, \\
b^{\rm HI}_{\rm IM}(z)&=&0.667 + 0.178 z + 0.050 z^2\,.
\eea

There is no threshold magnitude for IM, but there is a simple relation between the observed brightness temperature and the observed number counts of 21cm emitters, which allows us to use the number count contrast formulas \eqref{obdel} and \eqref{eq:fulltransfer}, with the IM clustering bias and the following evolution bias  and {\em effective} magnification bias \cite{Hall:2012wd,Fonseca:2015laa}:
\be \label{imsbe}
 (b_e)^{\rm HI}_{\rm IM}= { \p\ln \left[ \bar{T}^{\rm HI}{\cal H}\right]\over \p\ln a}-3\,,\qquad \big(s^{\rm HI}_{\rm IM}\big)_{\rm eff}={2\over 5}\,.
\ee 
Then we have
\be
\delta_T  = \delta_N \Big|_{b_e,s~ {\rm from~ \eqref{imsbe}}}\,.
\ee

\begin{table}
\caption{\label{tab:skadetails} SKA1 and MeerKAT: dish and receiver properties.}
\centering
\begin{tabular}{lcccc}
\\ \hline
Receiver & Frequency range & Redshift range &Dish size & \# dishes\\
&  [MHz]& & [m] & \\
\hline
Band 1& 350 -- 1050  & 0.35--3.06 &15 & 133\\
Band 2&  950 -- 1760& 0.1--0.49 &15 & 133\\
UHF-band& 580 -- 1015 & 0.4--1.45 &13.5 & ~64\\
L-band& 900 -- 1670 & 0.1--0.58 &13.5 & ~64\\
\hline
\end{tabular}
\end{table}

We consider  HI IM surveys with\footnote{www.skatelescope.org}  SKA1-Mid \cite{Bacon:2018dui}  and its precursor\footnote{www.ska.ac.za/science-engineering/meerkat/}  MeerKAT  \cite{Santos:2017aa}. MeerKAT is  operational; its 64 dishes will  eventually be incorporated into SKA1-Mid with 133 new dishes. 
Since MeerKAT will produce results earlier than SKA1, we consider it separately.
The total dishes, times and sky areas for the surveys are as follows: 
\bea
\mbox{SKA1 IM:}\quad &&N_{\rm d}=197\,,~  t_{\rm tot}=10,000\,{\rm hr}\,,~\Omega=20,000\,{\rm deg}^2 \,, \nn \\
\mbox{MeerKAT IM:}\quad && N_{\rm d}=~64\,,~ \, t_{\rm tot}=\,4,000\,{\rm hr}\,,~\,\Omega=~~4,000\,{\rm deg}^2 \,. \nn
\eea
Dish and receiver properties are shown in Table \ref{tab:skadetails}, where Band 1, 2 refer to the receivers of the 133 new dishes, while UHF- and L-bands refer to the receivers of the 64 MeerKAT dishes. We use the terms  SKA1 IM1,2 and MeerKAT IM-U,L for the high- and low-redshift surveys. The different properties of the MeerKAT and new SKA dishes complicate the thermal noise and beam properties of SKA1 IM1,2. The details are given in Appendix \ref{weight}.

\subsection{HI galaxy surveys}

For HI IM we considered radio dish arrays in single-dish mode. On the other hand, they can also be used as interferometers, increasing their angular resolution to be able to detect individual radio galaxies. If the 21cm line is present then one can measure the redshift of the galaxy, with extremely high redshift precision: this is the radio equivalent of an optical spectroscopic survey, with the noise also given by \eqref{eq:ngalbin}. 

We use the parametrisations of simulations of {HI galaxies} given in \cite{Yahya:2014yva} for  the redshift distribution and clustering bias:
\bea
N^{\rm HI}_g &\equiv& \frac{\d \mathbb{N}^{\rm HI}_g}{\d z \d\Omega}=10^{c_1} z^{c_2} \exp\big(-c_3 z\big) ~ {\rm gal}/\deg^{2}\,, \label{eq:nHIg} \\
b^{\rm HI}_g&=&c_4 \exp\big(c_5 z\big) \label{eq:bHIg}\,.
\eea
For the magnification bias we use the parametrisation from \cite{Camera:2014bwa}\footnote{Erratum MNRAS 467, 1505 (2017).}:
\be
s^{\rm HI}_g(z)=\frac25\left[ a_1(z)  + 2 a_2(z) \ln \frac{F_*}{\rm \mu Jy} + 3 a_3(z) \left(\ln \frac{F_*}{\rm \mu Jy}\right)^2\right]\,,
\ee
where $F_*$ is the threshold flux corresponding to $m_*$ in \eqref{eq:sg}.
The fit for the $a_i$ parameterisations can be found in table A1 of \cite{Camera:2014bwa} up to $z=1.5$, but has been computed\footnote{S. Camera, private communication.} up to $z=3$. The evolution bias is computed directly inside \cs using \eqref{eq:be} and \eqref{eq:nHIg}.
\begin{table}[!h]
\caption{\label{tab:HIgalpar} Fitting coefficients of  \eqref{eq:nHIg} and \eqref{eq:bHIg} for SKA HI  galaxy surveys.}
\centering
\begin{tabular}{lcccccc}
 \\ \hline
 &$c_1$ & $c_2$ & $c_3$ & $c_4$ & $c_5$ & $F_*$\\
\hline
SKA1 Gal & 5.45 & 1.31 & 14.4 & 0.616 & 1.02 &$ 100\mu$Jy \\
SKA2  Gal & 6.77 & 2.17 & 6.63 & 0.588 & 0.808 &$ 5\mu$Jy\\
\hline
\end{tabular}
\end{table}

We consider two HI galaxy surveys: the SKA1 Medium-Deep Band 2 Survey (hereafter SKA1 Gal) \cite{Bacon:2018dui} and  a more futuristic SKA2 Gal  \cite{Santos:2015hra}:
\bea
\mbox{SKA1 Gal:}\quad &&\Omega=~\,5,000\,{\rm deg}^2 \,,~~0.1\leq z \leq 0.58\,,~~F_*=100\,\mu{\rm Jy} \,, \nn \\
\mbox{SKA2 Gal:}\quad && \Omega=25,000\,{\rm deg}^2\,,  ~~0.1\leq z \leq  2.0~\, \,,~~F_*=~~\,5\,\mu{\rm Jy} \,. \nn
\eea
We assume that  $F_*=5\sigma$, where $\sigma$  is the flux sensitivity.
See Table \ref{tab:HIgalpar} for the values of the fitting coefficients in \eqref{eq:nHIg} and \eqref{eq:bHIg} for these two surveys {(using \cite{Yahya:2014yva})}.

\subsection{Optical/near-infrared spectroscopic surveys}

We consider one high- and one low-redshift future survey: 

\subsection*{Euclid-like  H$\alpha$ galaxy survey}

Euclid\footnote{www.euclid-ec.org} is a space telescope with an NISP spectrometer  in the near-infrared (NIR), $1100-2000\,$nm. Using the H$\alpha$ line to determine the redshift, this corresponds to $z=0.68-2.04$. 

Based on several datasets, \citep{Pozzetti:2016cch} have presented fits for H$\alpha$ luminosity functions as a function of redshift. From these results, and assuming that the flux threshold of an NISP-like spectrometer is  {$\sim2\times 10^{-16}$erg/s/cm$^{2}$}, \citep{Camera:2018jys} find the following fits for the redshift distribution of spectroscopic H$\alpha$ galaxies and their magnification bias:
\bea
N^{{\rm H} \alpha}_g  &\equiv& \frac{\d \mathbb{N}^{\rm H \alpha}_g}{\d z \d\Omega}= z^{1.11} \exp\big({9.91-0.84 z-1.26 z^2+0.40 z^3-0.04z^4\big)}~ {\rm gal}/\deg^2 \,, \label{eq:nHag} \\
s^{\rm H\alpha}_g&=& {0.49+0.41 z-0.17 z^2+0.05 z^3-0.01 z^4 \label{eq:bHag}\,,}\\
b^{\rm H\alpha}_{g} &=& \sqrt{1+z}\,.
\eea
The galaxy bias follows \citep{Amendola:2012ys,Amendola:2016saw}. 
\cs directly computes the evolution bias using \eqref{eq:be} and \eqref{eq:nHag}. 
We assume a  sky area of $15,000\,{\rm deg}^2$. 

\begin{table}[!h]
\caption{\label{tab:survdetails} Basic details of the surveys.}
\centering
\begin{tabular}{llrrrr}
\\ \hline
Experiment & Tracer &{Sky area}& $t_{\rm tot}$ & Redshift & Spectral\\
 & &$[10^3\deg^2]$& [$10^3$hr] & range & resolution\\
\hline
{MeerKAT IM-L}  & {HI IM}  & 4 & 4 & 0.1--{0.58}  &  {$10^{-4}$}\\
{MeerKAT IM-U}   &  {HI IM} & 4 & 4 & 0.4--1.45  &  {$10^{-4}$}\\
{SKA1 IM2}  & {HI IM}  & 20 & 10 & 0.1--0.58 & {$10^{-4}$}\\
{SKA1 IM1} & {HI IM} & 20 & 10 & 0.35--3.06 &  {$10^{-4}$}\\
{SKA1 Gal} & HI galaxies & 5 & --- & 0.1--0.58 &  {$10^{-4}$}\\
{SKA2 Gal} & HI galaxies & 25 & --- & 0.1--2.0  &  {$10^{-4}$}\\
Euclid-like & H$\alpha$ galaxies & 15 & --- &  0.68--2.05 & {$10^{-3}$}\\
DESI-like & Bright galaxies & 15 & --- & 0.1--0.6&  {$10^{-3}$}\\
\hline
\end{tabular}
\end{table}

\subsection*{DESI-like bright galaxy sample}  

The Dark Energy Spectroscopic Instrument (DESI)\footnote{www.desi.lbl.gov} is a ground-based experiment whose spectrometer will determine the redshift of millions of galaxies with a redshift resolution of around 1$\%$. The survey will target, among others, Luminous Red Galaxies (LRG), Emission Line Galaxies (ELG), Quasars and a Bright Galaxy Sample (BGS) \cite{Aghamousa:2016zmz}. The BGS sample is at low redshifts ($z=0.05-0.6$) while the others are at higher redshifts. Here we consider a DESI-like BGS survey only, as a low-redshift complement to a Euclid-like H$\alpha$ survey -- which on its own provides better constraints than the higher-redshift samples of a DESI-like survey. 

We use the fits to simulations from \cite{Aghamousa:2016zmz} for the astrophysical properties of these galaxies:
\bea
N^{\rm BGS}_g &\equiv&  \frac{\d \mathbb{N}^{\rm BGS}_g}{\d z \d\Omega}= 6.0\times 10^3 \l \frac z{0.28}\r^{0.91} \exp\Big[-\Big({z\over 0.28}\Big)^{2.56}\Big] ~ {\rm gal}/\deg^{2}\,, \label{eq:nbgs} \\
b^{\rm BGS}_g&=&0.99+0.73 z-1.29 z^2+10.21 z^3  \label{eq:bbgs}\,.
\eea
The evolution bias is computed directly inside \cs using \eqref{eq:be} and \eqref{eq:nbgs}. We neglect the lensing contribution to the angular power spectrum, which is a very good approximation for low-redshift spectroscopic surveys.

\paragraph{}  \hspace{0pt} \\
We summarise in Table \ref{tab:survdetails} the basic observational details of the surveys considered.

\section{{Results}}
\subsection{Redshift binning and the precision on RSD} \label{sec:optbin}

\begin{figure}[]
\centering
\vspace*{-0.5cm}
\includegraphics[width=15cm]{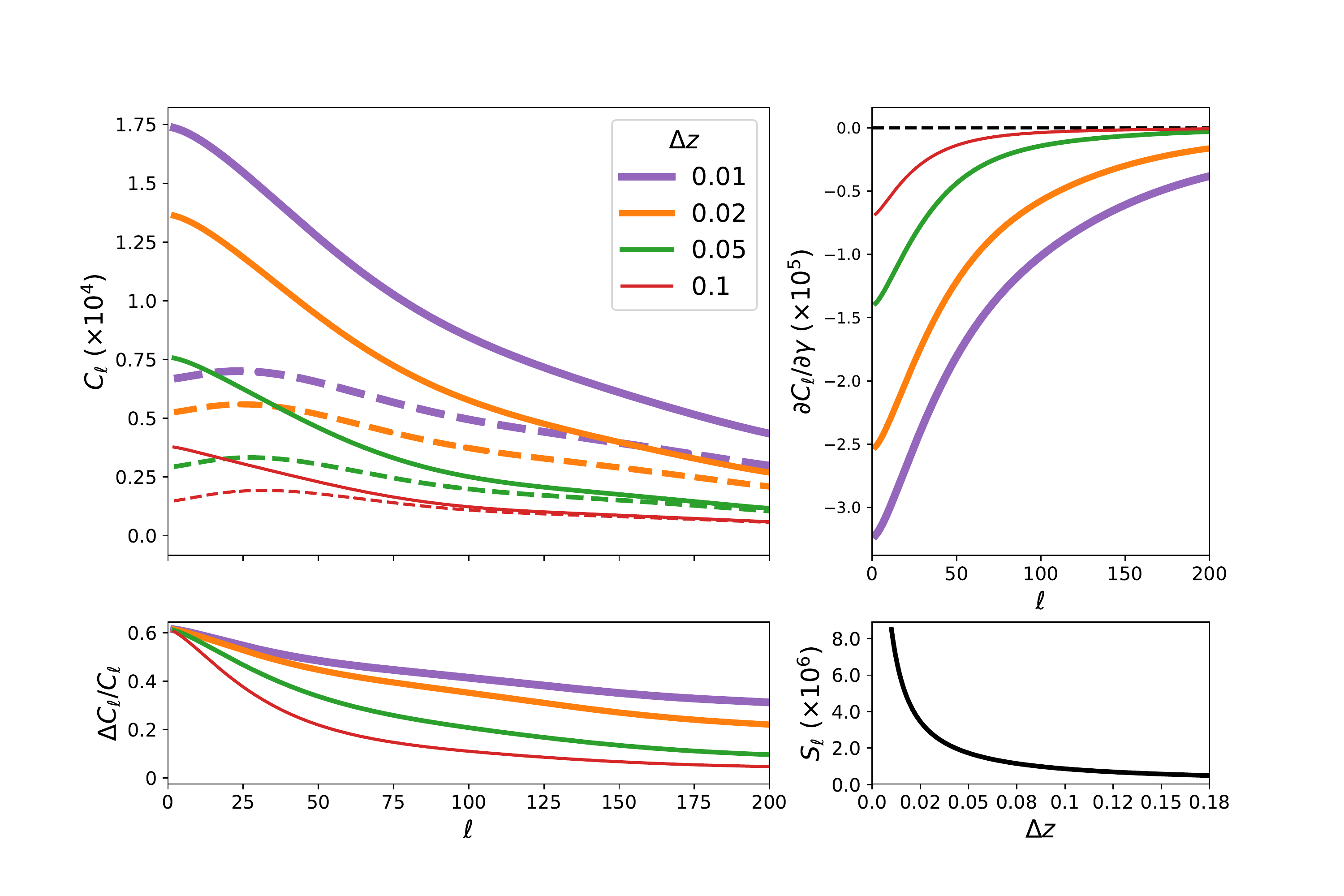}\\
\caption{For a Euclid-like survey, with different bin widths at $z_i=1$. {\em Top left:}  $C_\ell(z_i,z_i)$ without RSD (dashed) and with RSD (solid). {\em {Bottom} left:} fractional RSD contribution to total angular power, where $\Delta C_\ell=C_\ell({\rm total})-C_\ell({\rm density\ only})$. {\em Top right:} Fisher derivative $\partial C_\ell/\partial \gamma$. {\em {Bottom} right:} shot noise \eqref{slg}.}
\label{rsdcl}
\end{figure}
In a Fisher analysis of the Fourier power 
spectrum  $P_g({\k})$, the redshift bin width does not affect the analysis, provided that $\Delta z$ is not too large (typically $\Delta z \lesssim 0.1$ is chosen). The important binning in the Fourier case is in $k$-space. For the angular power spectrum $C_\ell$, the Fisher analysis is different and the bin width is important for spectroscopic surveys. In this case, we need to apply redshift binning in a way that corresponds to the observable of interest. For example, if we choose broad redshift bins, then this will suppress the RSD signal, since peculiar motions will tend to average out.
We can confirm this expectation by looking at the relative strength of RSD in the angular power spectrum as the bin width is decreased. A typical example is shown in Fig. \ref{rsdcl} (top left), where we plot $C_\ell(z_i,z_i)$ without RSD (dashed) and with RSD (solid), for different bin widths $\Delta z$; the fractional RSD contribution is shown in the {bottom} left panel. The Fisher information on the growth rate index $\gamma$ also increases with decreasing bin width, as shown in the top right plot.

It is clear from these plots that one can in principle extract more information from RSD by decreasing the width of the redshift bin. On the other hand, this also  increases the shot  noise \eqref{slg}, since $N_\Omega \sim N \Delta z$ (and similarly for the instrumental noise \eqref{eq:inst_noise_hiim}). The increase in shot noise with decreasing bin width is illustrated in Fig. \ref{rsdcl} ({bottom} right). 

\begin{figure}[]
\centering
\vspace*{-0.5cm} 
\includegraphics[width=7.65cm]{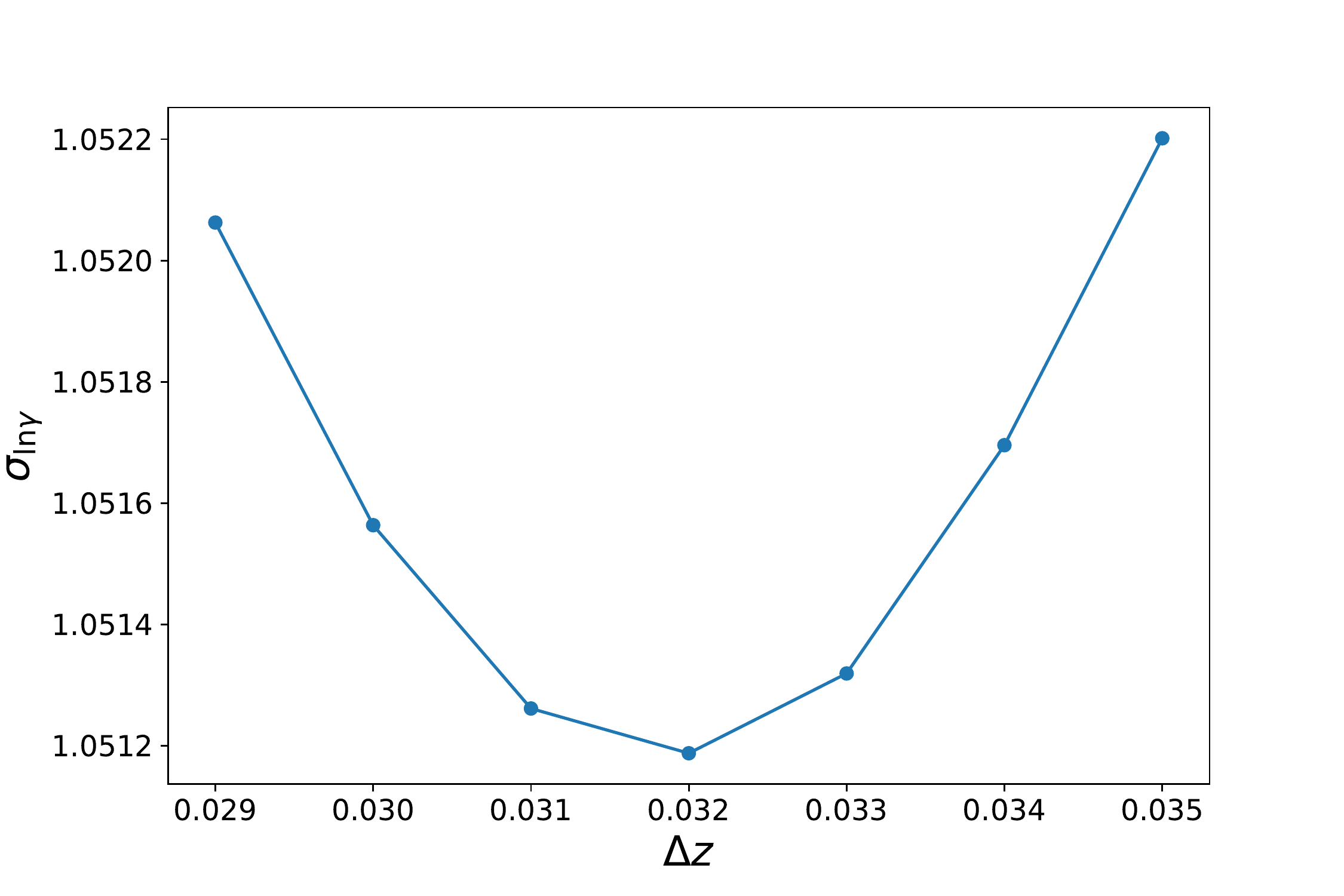}
\includegraphics[width=7.65cm]{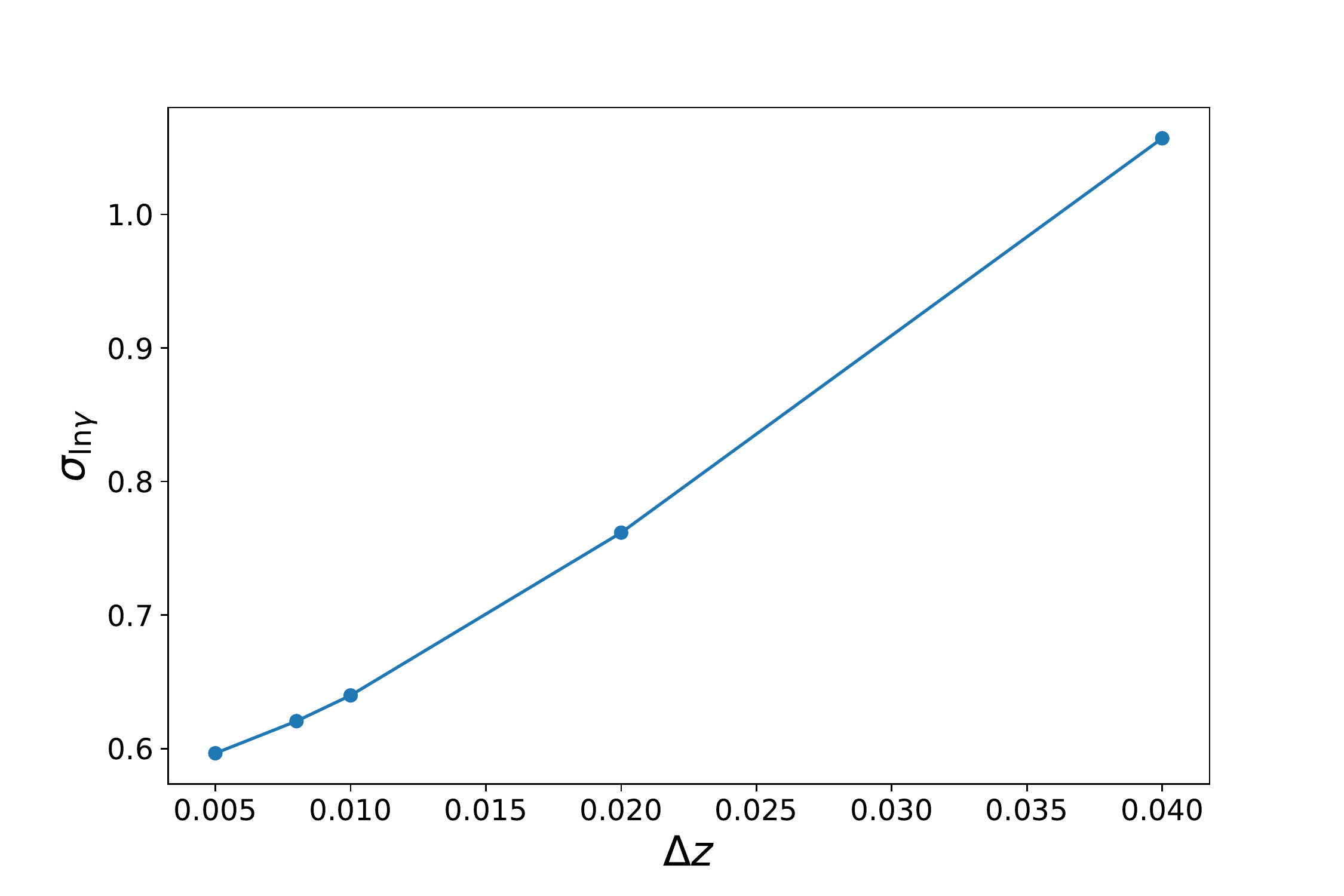}
\caption{SKA1 IM1 survey centred at $z_i=2$. Dependence of the conditional $\gamma$ error on the redshift bin width:   for a single bin of varying width ({\em left}); for a growing number of sub-divisions with fixed total size of 0.04, including all  auto- and cross-correlations ({\em right}).}
\label{fig:min_and_xcor}
\end{figure}

One might expect that the battle between increasing signal and growing noise produces a `sweet spot' where information is maximised. The left panel of Fig. \ref{fig:min_and_xcor} appears to show this, for a bin centred at redshift $z_i=2$ in a SKA1 IM1 survey. We vary the width of a single redshift bin and  find an optimal bin size that minimises the conditional error from auto-correlations. However, the problem with this apparent optimisation, is that we have neglected {\em cross-correlations}. Decreasing the bin width means an increase in the number of bins, and consequently an even greater increase in the number of auto- and cross-spectra amongst the redshift bins. The additional information from these correlations compensates for the increased noise per bin. This is illustrated in the right panel of Fig. \ref{fig:min_and_xcor}. We fix the total redshift range to 0.04, subdivide it into bins 8 bins of $\Delta z=0.005$, 5 bins of $\Delta z=0.01$, 2 bins with $\Delta z=0.02$ and a single bin with $\Delta z=0.04$. Then we compute the conditional constraints on $\gamma$ including cross-correlations between redshift bins. The `sweet spot' at $\Delta z\sim 0.032$ in the left panel is not seen once the cross-correlations are included. Therefore we can in principle reduce the bin size down to the size of the receiver bands, which is $\sim 3\times 10^{-4}$  at $z_i=2$.

The implication is that we should choose a redshift bin width that is as small as possible, given the practical constraints imposed by redshift resolution and especially by numerical computation. In order to extract all the information, we need to include cross-correlations between bins -- and this becomes numerically prohibitive for very large numbers of bins.
 To tackle this problem, we follow the `hybrid' method proposed in \cite{Camera:2018jys} -- i.e., we divide the full redshift range of a survey into sub-surveys and perform all cross-bin correlations in each sub-survey, but not between sub-survey bins. There is a small loss of information from neglecting some cross-correlations between redshift bins. However,  if the sub-surveys are wide enough, i.e., bigger than the correlation length (which is typically $\sim 0.1$ in redshift), this loss is only non-negligible for adjacent bins of sub-surveys  \cite{Camera:2018jys}. We use a redshift bin-width of $\Delta z=0.01$, which is numerically feasible with the hybrid method.

In the hybrid approach, the constraints from a survey are just the summed constraints from each  sub-survey. We modify this slightly in order to deal with the survey-dependence of the clustering  bias. We marginalise over the bias parameters $b(z_i)$ in  the  Fisher matrices for each sub-survey, before adding these matrices to obtain the Fisher matrix of the full survey. In more detail: 
let $^{I}\tilde F_{\varphi_\alpha \varphi_\beta}$ be the Fisher matrix of sub-survey $I$ marginalised over the clustering bias, so that $\varphi_\alpha$ are all the parameters in  \eqref{para} except for $b^{ I}(z_i)$. Then
\be
^{I}\tilde F_{\varphi_\alpha \varphi_\beta}= \left[\l {}^{I}F_{\vartheta_\alpha \vartheta_\beta}\r^{-1}_{\varphi_\alpha \varphi_\beta}\right]^{-1}\,,
\ee 
and the total Fisher matrix is 
\be
\tilde F_{\varphi_\alpha \varphi_\beta}=\sum_{I} {}^{I}\tilde F_{\varphi_\alpha \varphi_\beta}\,.
\ee

In the case of the low-redshift surveys BGS, SKA1 IM2, SKA1 Gal, and MK IM-L, the number of bins is low enough compute the full   tomographic result. We compared this with the result from the sub-survey approximation with 2 sub-surveys, and found that the sub-survey approximation is only slightly worse, at the second significant figure.

\subsection{Forecast results}\label{sec:results}

\begin{figure}[!h]
\centering
\vspace*{-0.3cm}
\includegraphics[width=13cm]{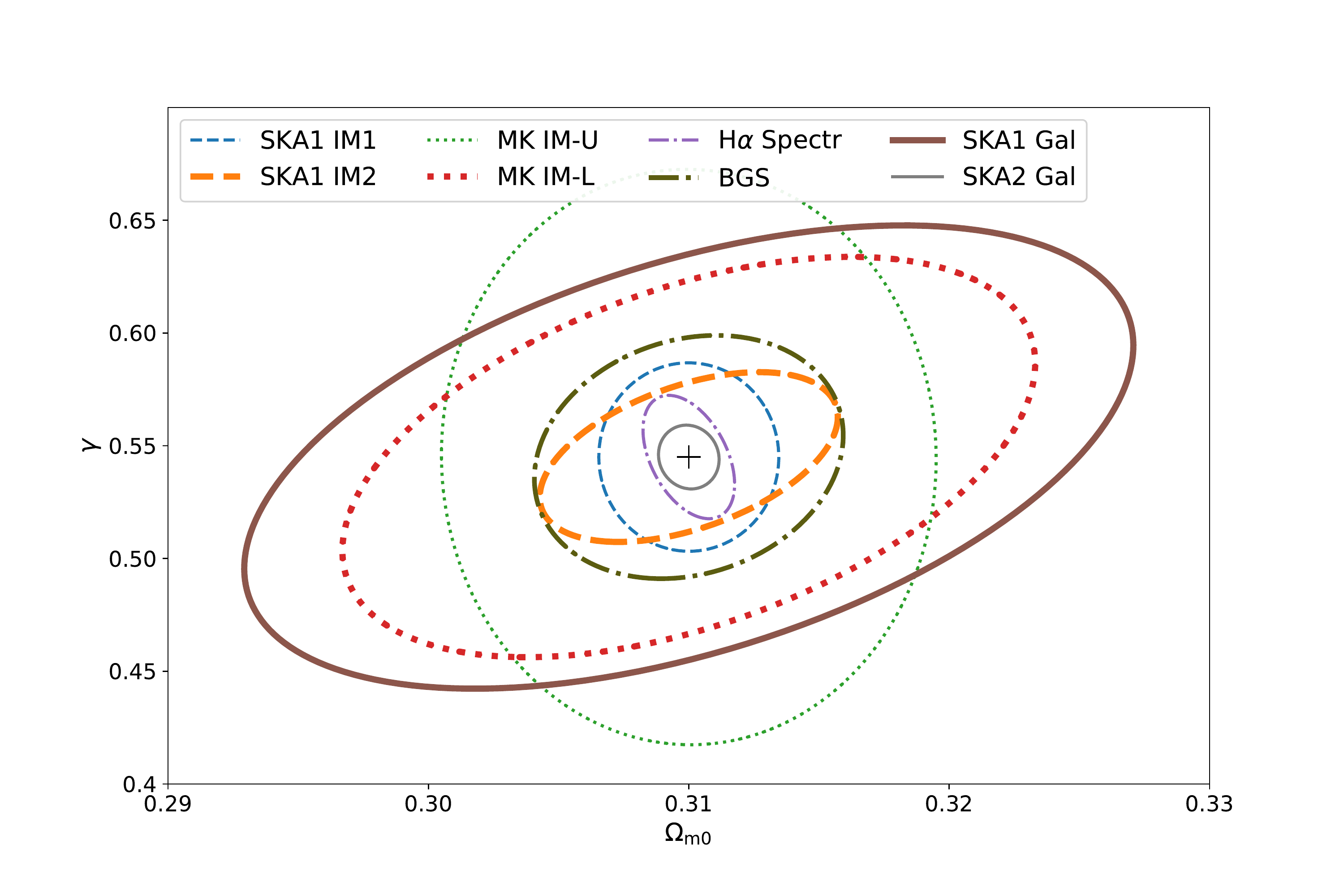}
\caption{ Marginal {$1\sigma$}  contour plots of $\gamma$ and {$\Omega_{\rm m0}$} for the surveys considered. Plus sign indicates the fiducial values. No priors included.}
\label{fig:contour_errors}
\end{figure}

We summarise the errors on $\gamma$ in Table \ref{tab:marginal_gamma}. The best forecasts (including Planck 2018 priors) are in the range $\sim3-5\%$ for the near-future surveys. This is only  improved to percent level in the more futuristic HI galaxy survey with SKA2. In Figure \ref{fig:contour_errors} we show the contour plots for $\gamma$ and the total matter density today, $\Omega_{\rm m0}=\Omega_{\rm cdm0}+\Omega_{\rm b0}$. 
\begin{table}[!h]
\caption{\label{tab:marginal_gamma} Errors on $\gamma$ in spectroscopic surveys, with and without Planck 2018 priors. Note that the last sub-survey may have a different number of bins as it is just the remainder.}
\centering
\begin{tabular}{llrrrrr}
\\ \hline
 Redshift &Survey  & \multicolumn{2}{p{0cm}}{Subsurveys}  &  {\centering $\sigma^{\rm cond}_{\ln \gamma}$} & \multicolumn{2}{p{3cm}}{\centering $\sigma_{\ln \gamma}$}\\
& & $\#$  & $\#$ bins  & & no prior & with prior\\
& &  &  {each}  &{\%}  &{\%} &{\%}\\
\hline
Low  redshift &BGS & 1 & 50 & 2.6& 6.7& 4.5\\
&SKA1 Gal & 1 & {48} & {3.6} &  {14.0} & {6.4}\\ 
&{MK IM-L} & 1 & {48} &  {2.9} &  {10.8} &  {5.7}\\
&SKA1 IM2 & {1} & {48} &  {1.2} &  {4.6} & {2.8}\\
\hline
High redshift &H$\alpha$ Spectr &  {3} &  {45} &  {1.4} &  {3.9}  &   {2.9} \\ 
&{MK IM-U} &  {2} &  {52} &  {2.9} &  {17.8} &  {8.0}\\
&SKA1 IM1 &  {5} &  {45} & {1.2} &  {5.3} &  {3.7}\\
\hline
Low and high&SKA2  {Gal} &  {4} & {47} &  {0.6} &  {1.7} &  {1.4}\\
\hline
\end{tabular}
\end{table}
%

\section{Conclusions} \label{sec:conclusion}

In this paper, we computed constraints on the growth index $\gamma$ using the observed power spectrum, i.e. the angular power spectrum $C_\ell$, of spectroscopic cosmological surveys. Instead of an exhaustive study of surveys, we tried to fill the redshift range $0<z \lesssim 3$ with the potentially best contemporaneous spectroscopic surveys. We first investigated the effect of redshift bin width  on the amount of information on $\gamma$ that can be extracted. Unlike the case of the Fourier power spectrum  $P_g$, the choice of redshift bin width has a significant impact in the angular power spectrum. We showed that cross-correlations between bins compensates for the growth in noise, and we concluded that in theory, the thinnest possible width will deliver the highest signal-to-noise. Numerical constraints from the high number of cross-correlations between redshift bins can  be mitigated by a  `hybrid' method, and we chose a bin width of $\Delta z=0.01$.

Key advantages of $C_\ell$ include: it incorporates the redshift evolution of all cosmological, astrophysical and noise variables; it does not impose a flat-sky approximation but naturally incorporates wide-angle correlations; Doppler and lensing corrections to the 2-point correlations are also naturally included. Furthermore, since it is directly observable, the angular power spectrum of the data requires no fiducial model and therefore no Alcock-Paczynski correction is needed. 

\begin{table}[]
\caption{\label{nocc} As in Table \ref{tab:marginal_gamma}, but {\em neglecting} all cross-bin correlations}
\centering
\begin{tabular}{llrrrr}
\\ \hline
 Redshift &Survey   &  {\centering $\sigma^{\rm cond}_{\ln \gamma}$} & \multicolumn{2}{p{3cm}}{\centering $\sigma_{\ln \gamma}$}\\
&  & & no prior & with prior\\
&  &{\%}  &{\%} &{\%}\\
\hline
Low  redshift &BGS &  {3.6}&  {26.1}&  {8.4}\\
&SKA1 Gal  &  {4.4} &  {62.0} &  {9.8}\\ 
&{MK IM-L} &  {3.3} &  {65.2} &  {9.1}\\
&SKA1 IM2  &  {1.4} & {27.4} &  {5.1}\\
\hline
High redshift &H$\alpha$ Spectr &  {1.7} &  {14.6} &  {7.2} \\ 
&{MK IM-U} &  {3.2} &  {124.2} &  {11.0}\\
&SKA1 IM1 & {1.3} &  {26.7} &  {6.4}\\
\hline
Low and high&SKA2  Gal &  {0.7} &  {4.9} &  {3.4}\\
\hline
\end{tabular}
\end{table}
These advantages over the Fourier power spectrum $P_g$ (which is not an observable) come with a price. Unlike $P_g$,  $C_\ell$ does not allow a clean separation of the RSD effect.  In addition, there are computational challenges in extracting maximal information from  $C_\ell$. In particular, performing all cross-bin correlations becomes increasingly difficult for the very thin bins. We used a variant of a `hybrid' method to capture the dominant cross-correlation contribution. Including cross-bin contributions is very important. In Table \ref{nocc} we show the constraints computed when neglecting all cross-bin correlations and using only auto-correlations. By comparing with Table \ref{tab:marginal_gamma}, we see that the marginal constraints with priors are degraded by a factor of $\sim 40-150\%$, and those without priors are degraded by much more.

In our Fisher forecasts, we marginalised over the standard cosmological parameters, as well as the dark energy equation of state and the clustering bias in each redshift bin, for each survey. Our constraints are based only on the information from linear scales.
Our main results are shown in Table \ref{tab:marginal_gamma} and in the error contour plots of Figure \ref{fig:contour_errors}. The best marginal constraints (including priors) on $\gamma$ are $\sim3-5\%$ for the near-future surveys, with SKA1 intensity mapping providing the best near-future constraints, while the SKA precursor MeerKAT is predicted to be competitive. The more futuristic SKA2 HI galaxy survey should reach sub-percent errors. 

\[\]
\acknowledgments
We thank an anonymous reviewer for alerting us to a key point, and Stefano Camera, Sheean Jolicoeur, Alkistis Pourtsidou and M\'ario G. Santos for useful comments and discussions. JF is supported by the University of Padova under the STARS Grants programme CoGITO: Cosmology beyond Gaussianity, Inference, Theory and Observations. 
JV and RM are supported by the South African Radio Astronomy Observatory and the National Research Foundation (Grant No. 75415). RM is also supported by the UK Science \& Technology Facilities Council (Grant ST/N000668/1).

\newpage
\appendix

\section{SKA1 intensity mapping noise}\label{weight}

We need to weight the noise and beam from the 64 MeerKAT  dishes and the 133 new dishes, with different diameters and  receiver bands (see Table \ref{tab:skadetails}).   We weight the contributions of the different antennas using the individual RMS. For the UHF/L-bands and Band 1/2 we have:
\bea
w_{\rm U/L} =w_{\rm tot}^{-1}\ \frac{N_{\rm d,U/L}}{T^2_{\rm sys, U/L}}\,,\quad
w_{\rm 1/2} = w_{\rm tot}^{-1}\ \frac{N_{d,\rm 1/2}}{T^2_{\rm sys, 1/2}}\,,  
\eea
where
\bea
w_{\rm tot} = \frac{N_{ \rm d,U/L}}{T^2_{\rm sys, U/L}}+\frac{N_{ \rm d,1/2}}{T^2_{\rm sys, 1/2}}\,.
\eea
Then the weighted instrumental noise for SKA1  surveys is given by
\be \label{wis}
{\cal T}_\ell^{\rm SKA1\,  IM1/2}=w_{\rm  U/L} {\cal T}_\ell^{\rm  U/L} + w_{\rm 1/2} {\cal T}_\ell^{\rm 1/2}=\frac{2\pi\ f_{\rm sky}}{\Delta \nu\ t_{\rm tot}}\frac{T_{\rm sys,eff}^2}{N_{\rm d,tot}}\,,
\ee
where  ${\cal T}_\ell^{\rm  U/L}$ is the noise for 64 MeerKAT dishes in UHF/L-bands, and ${\cal T}_\ell^{\rm  U1/2}$ is the noises for 133 new SKA dishes in Bands 1/2 -- each 
given by \eqref{eq:inst_noise_hiim}. 

\begin{figure}[!h]
\includegraphics[width=8.2cm]{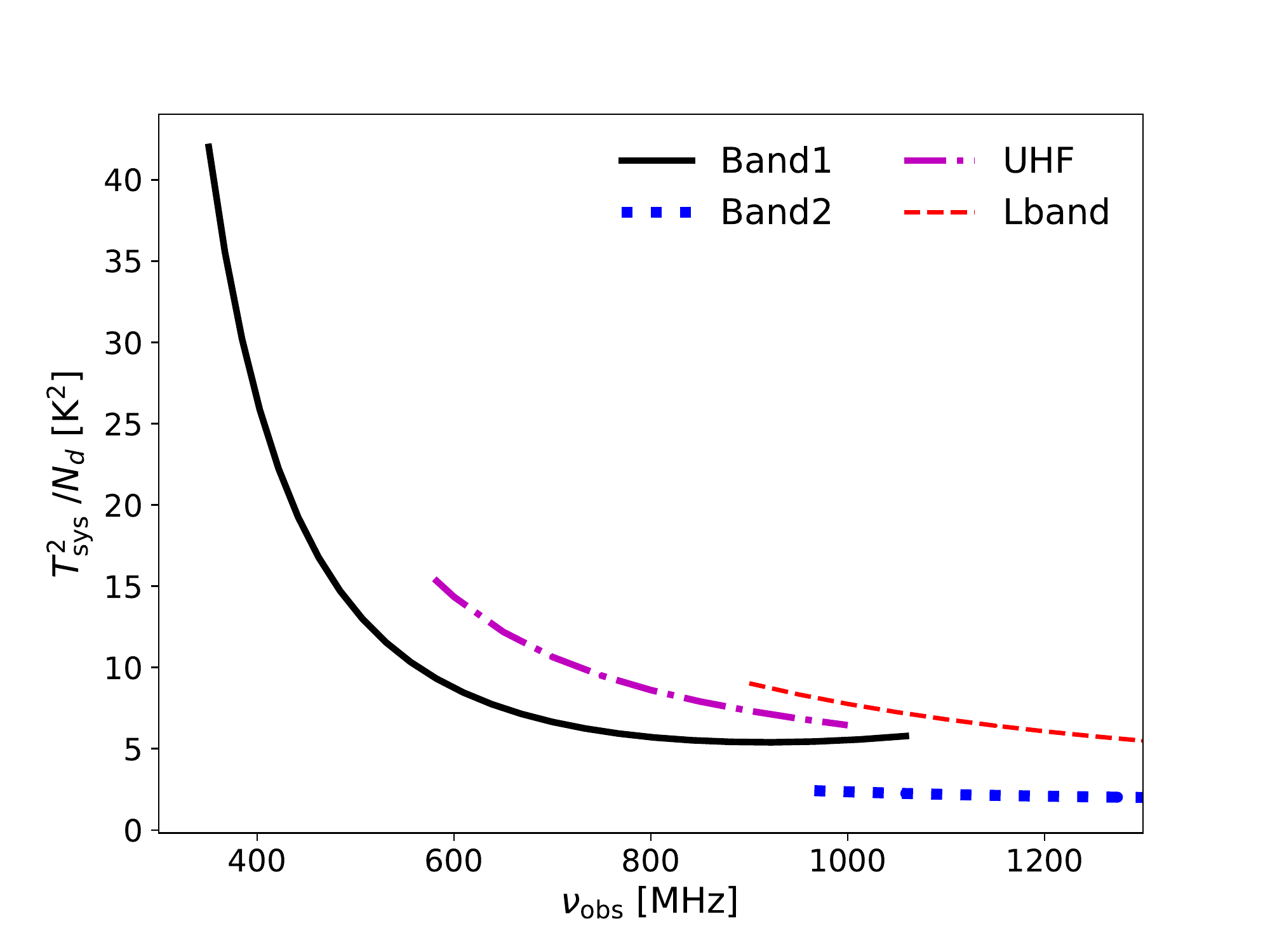}
\includegraphics[width=8.2cm]{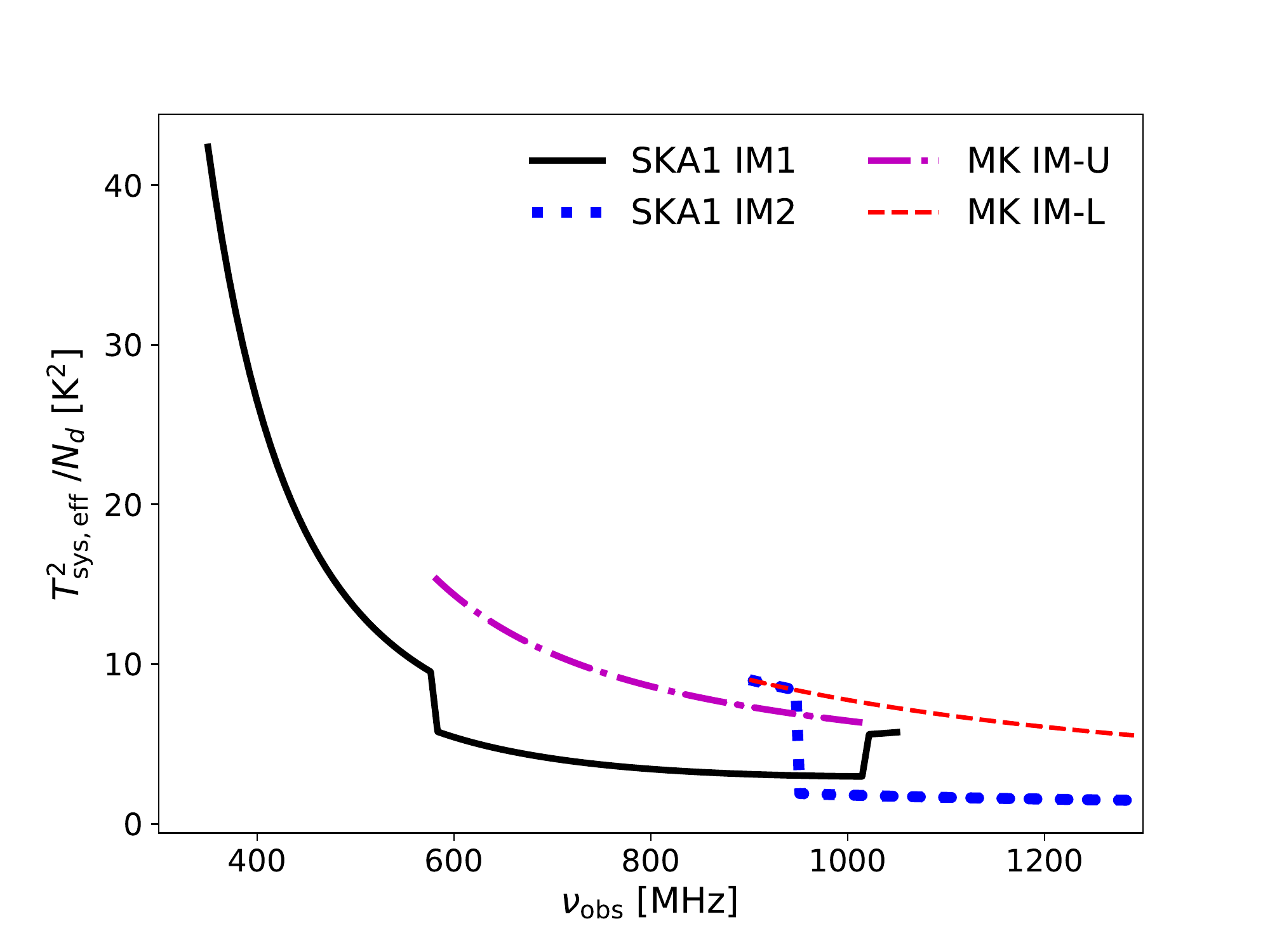}
\caption{  System temperature (in the form $T^2/N$) for the 4 receiver bands of HI IM (\emph{left}) and the weighted effective system temperature for SKA1 and MeerKAT (\emph{right}). }
\label{fig:tsyseff}
\end{figure}

The system temperatures, in the form $T^2/N$, are shown in Figure \ref{fig:tsyseff}, for the individual receivers (left) and for SKA1 using the weighted noise \eqref{wis} (right).
For the MeerKAT bands the system temperature and the {effective} one are the same. Note that the jumps in the effective system temperature for SKA1 IM1/2 arise from the fact that the frequency range of the SKA1 bands and the MeerKAT bands do not perfectly overlap.

Finally, the total weighted beam is simply given by
\be
B_\ell^{\rm SKA1\, IM1/2}=w_{\rm 1/2} B_\ell^{\rm 1/2}+ w_{\rm U/L} B_\ell^{\rm U/L}\,.
\ee

\newpage

\providecommand{\href}[2]{#2}\begingroup\raggedright

\endgroup

\end{document}